\def\aj{{AJ}}

\def\apj{{ApJ}}
\def\apjs{{ApJS}}

\def\mnras{{MNRAS}}

\def\etal {{et al.~}}
\def\ddeg{{\rlap.}$^{\circ}$}

\documentclass[cup5b]{caps}
\usepackage{graphicx}
\usepackage{amssymb}
\usepackage{ociwsymp2}   

\HeadText{L. A. Page}

\begin{document}

\pagenumbering{arabic}

\author[]{LYMAN A. PAGE\\Department of Physics, Princeton University}

\chapter{The Wilkinson Microwave \\ Anisotropy Probe}

\begin{abstract}

The {\it Wilkinson Microwave Anisotropy Probe (WMAP)}\ has mapped the
full sky in five frequency bands between 23 and 94 GHz.
The primary goal of the mission is to produce high-fidelity, 
all-sky, polarization-sensitive maps that can be used to study the cosmic 
microwave background.
Systematic errors in the maps are constrained to new levels: using
all-sky data, aspects of the anisotropy may confidently be probed
to sub-$\mu$K levels. 
We give a brief description of the instrument and an overview of
the first results from an analysis of maps made with one year of data.
The highlights are (1) the flat $\Lambda$CDM model
fits the data remarkably well, whereas an Einstein-de~Sitter model does not; 
(2) we see evidence of the birth of the
first generation of stars at $z_r\approx 20$; (3) when the {\sl WMAP} data 
are combined and compared with other cosmological probes, a cosmic 
consistency emerges: multiple different lines of inquiry lead to
the same results. 

\end{abstract}

\section{Introduction}

We present the first results from the {\it Wilkinson Microwave Anisotropy 
Probe (WMAP)}, which completed its first year of observations
on August 9, 2002. Dave Wilkinson, for whom the satellite was renamed,
was a friend and colleague to many of the conference participants.
He was a leader in the development of the cosmic
microwave background (CMB) as a potent cosmological
probe. He died on September 5, 2002, after battling cancer for 17 years,
all the while advancing our understanding of the origin and evolution 
of the Universe. He saw the data in their full glory and was pleased
with what I am able to report to you in the following. 

At the time of the conference, the data were not ready to be released. 
The presentation focused on what {\sl WMAP} would add to the 
considerable advances that had been made over the past few years.
The emphasis was on the data; a companion talk by 
Ned Wright
showed how the data are interpreted in the context of the currently
favored models. Some effort was made to show what we knew firmly, like
the position of the first acoustic peak in the CMB, as opposed
to quantities that we knew better through the combination of multiple
experiments, such as the characteristics of the second acoustic peak 
and the possible level of foreground contamination.

Since the conference, there have been 17 papers by the {\sl WMAP}
team\footnote{Science team members are C. Barnes, C. Bennett (PI),
M. Halpern, R. Hill, G. Hinshaw, N. Jarosik, A. Kogut, 
E. Komatsu, M. Limon, S. Meyer, N. Odegard, L. Page, 
H. Peiris, D. Spergel, G. Tucker, L. Verde, J. Weiland, 
E. Wollack, and E. Wright.}
describing the instrument and data. This contribution draws 
heavily from those papers. 
Bennett \etal~(2003a) lay out the goals of the mission and give
an overview of the instrument. More detailed descriptions of
the radiometers, optics, and feed horns are found in 
Jarosik~\etal~(2003a), Page~\etal~(2003a), and Barnes~\etal~(2002). 
The inflight 
characterization of the receivers, optics, and sidelobes are 
described in  Jarosik~\etal~(2003b), Page~\etal~(2003b), 
and Barnes~\etal~(2003).
The data processing and checks for systematic errors
are described by Hinshaw~\etal~(2003a), and 
Hinshaw~\etal~(2003b) 
present the angular power spectrum. Bennett~\etal~(2003c) present the 
analysis of the foreground emission.
The temperature-polarization correlation is described
in Kogut~\etal~(2003), and Komatsu~\etal~(2003) 
show that the fluctuations are Gaussian.
Verde~\etal~(2003) present the details of the analysis of the angular power
spectrum and show how the {\sl WMAP} data can be combined with external data 
sets such as the 2dFGRS (Colless~\etal~2001). Spergel~\etal~(2003) present 
the cosmological parameters derived
from the {\sl WMAP} alone, as well as the parameters derived from
{\sl WMAP} in combination with the external data sets.
Peiris~\etal~(2003) confront the data with models of the early Universe, 
in particular inflation. The features in the power spectrum are interpreted
in Page~\etal~(2003c). The scientific results are summarized in 
Bennett~\etal~(2003b).

\begin{table}
%\centering
\caption{Characteristics of the Instrument}
\begin{tabular}{c|ccccc}
\hline \hline
Band & $\nu_{center}$ & $\Delta \nu_{noise}$ & $N_{chan}$
& $\Delta T/pix$ & $\theta_{FWHM}$ \\
     & (GHz)          & (GHz)          &  & ($\mu$K) & (deg)\\
\hline
       K  & 23 & 5  & 2 & 20  & 0.82 \\
       Ka & 33 & 7  & 2 & 20  & 0.62 \\
       Q  & 41 & 7  & 4 & 21  & 0.49 \\
       V  & 61 & 11 & 4 & 24  & 0.33 \\
       W  & 94 & 17 & 8 & 23  & 0.21 \\
\hline \hline
\end{tabular}
\label{tab:inst}

\footnotesize{
The $\Delta T/pix$ is the Rayleigh-Jeans sensitivity per $3.2\times10^{-5}$
sr pixel for \\ the four-year mission for all channels of
one frequency combined.}
\end{table}

\begin{figure*}[t]
\includegraphics[width=1.0\columnwidth,angle=0,clip]{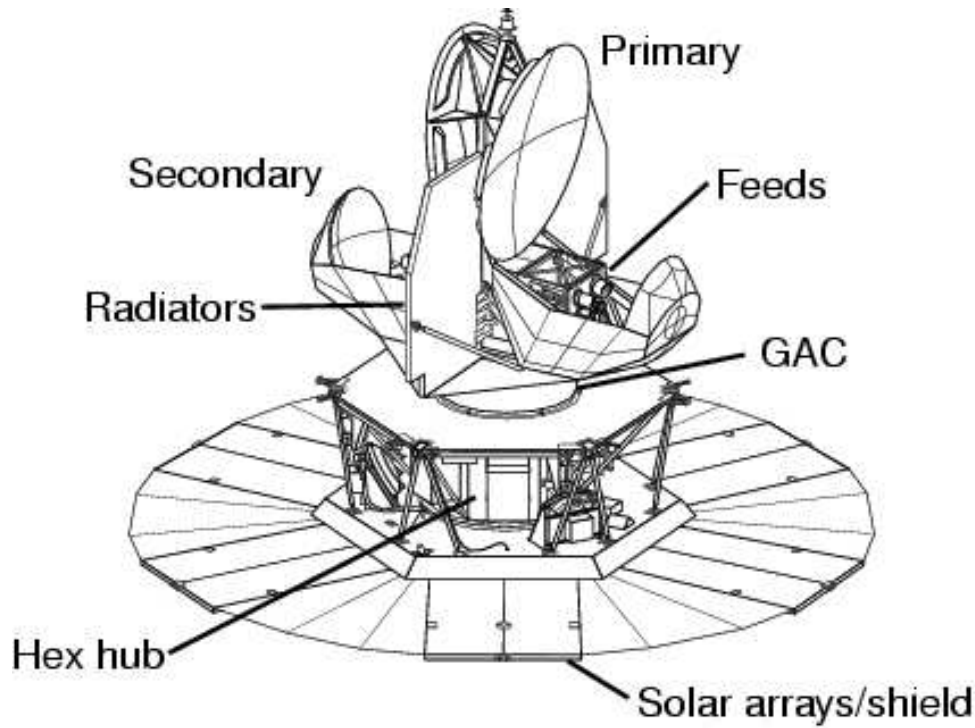}
\vskip 0pt \caption{
Outline of the {\sl WMAP} satellite. The overall height is
3.6~m, the mass is 830 kg, and the diameter of the large disk on
the bottom is 5.1 m. Six solar arrays on the bottom of this disk
supply the 400~W to power the spacecraft and instrument.
Thermal blanketing between the hex hub and GAC, and
between the GAC and radiators, shield the instrument from thermal
radiation from the support electronics and attitude control systems.
\label{fig:map_line}}
\end{figure*}

\begin{figure*}[t]
\includegraphics[width=1.0\columnwidth,angle=0,clip]{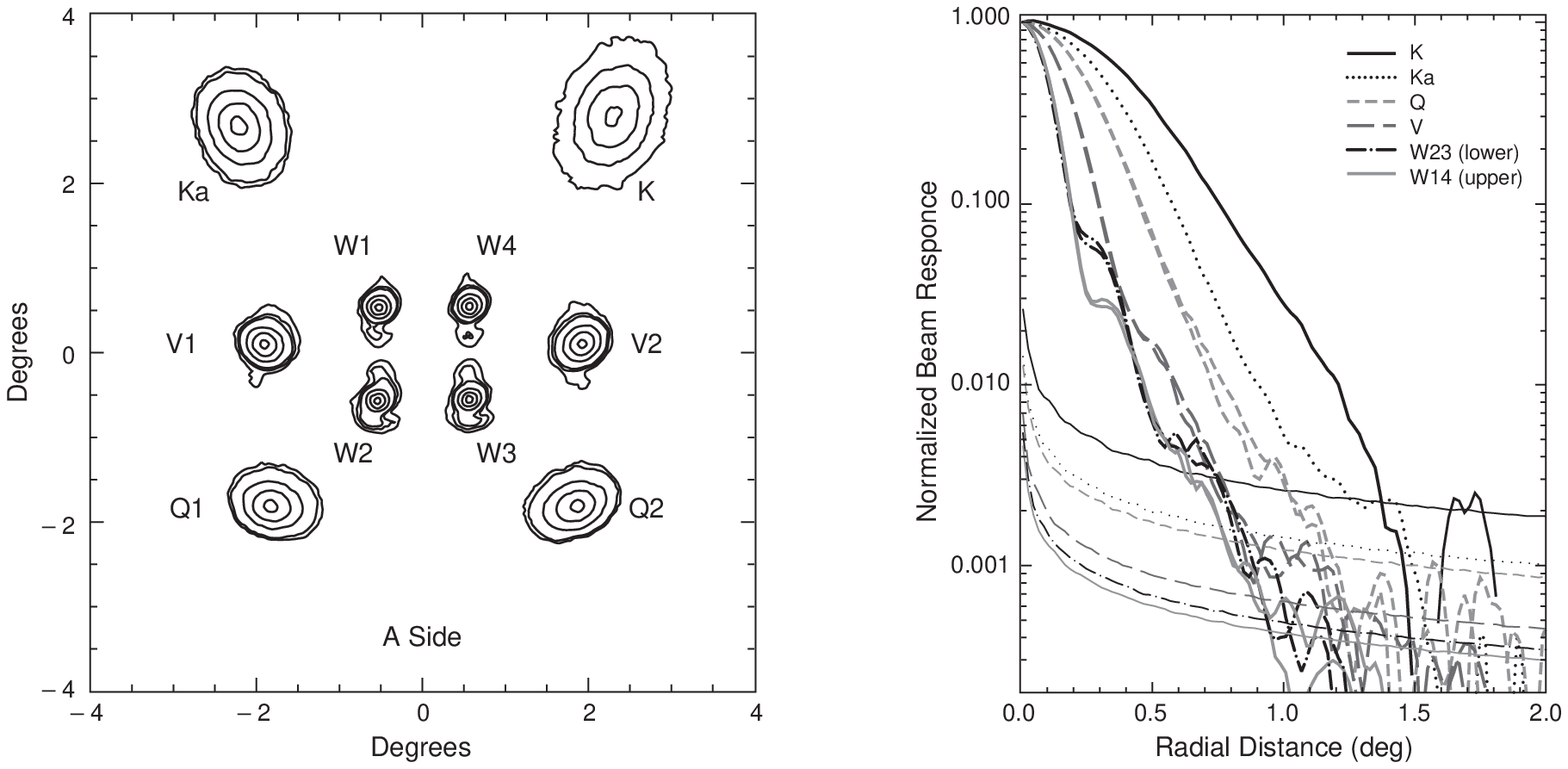}
\vskip 0pt \caption{
{\it Left:} The A-side focal plane obtained from observations
of Jupiter. Note that the beam widths
are a function of frequency. The farther a beam is from the central
focus, the more ellipsoidal it is. In the V and W bands, the beams
have substructure. This is due to deformations in the surface of the
primary and was anticipated. The scan pattern symmetrizes all the beams.
The contour levels correspond to 0.9, 0.6, 0.3, 0.09, etc. of the peak value.
{\it Right:} The normalized, symmetrized
beam response. Except in the K band, the beams are mapped to better than
$-30$~dB of their peak value. The lower set of lines shows the noise
level in the maps. From Page~\etal~(2003b).
\label{fig:focal}}
\end{figure*}

\section{Overview of the Mission and Instrument}

The primary goal of {\sl WMAP} is to produce high-fidelity, 
polarization-sensitive maps of the microwave sky that may be
used for cosmological tests.
{\sl WMAP} was proposed in June 1995, at the height of the ``faster,
better, cheaper'' era; some building began in June 1996, although the major
push started in June 1997 after the confirmation review. {\sl WMAP}
was designed to be 
robust, thermally and mechanically stable, built of components
with space heritage,\footnote{Other than the NRAO HEMT-based (High Electron
Mobility Transistors) amplifiers, which were
custom designed, all components were ``off-the-shelf.''} 
and relatively easy to integrate and test. 

There is no {\it single} thing on {\sl WMAP} that sets the mission
apart from other experiments. Rather, it is a combination 
of attributes that are designed to work in concert to
take full advantage of the space environment. The mission goal
is to measure the microwave sky on angular scales of $180^\circ$
as well as on scales of 0\ddeg2 with micro-Kelvin {\it accuracy}.
During the planning, building, and integration of the satellite,
the emphasis was on minimizing systematic
errors so that the data could be straightforwardly interpreted.
There are many levels of redundancy that permit multiple
cross-checks of the quality of the data. 

{\sl WMAP} was launched on June 30, 2001 into an orbit 
that took it to the second 
Earth-Sun Lagrange point, $L_2$, roughly $1.5\times10^6~$km 
from Earth. The $L_2$ orbit is oriented so that the instrument is always
shielded from the Sun, Earth, and Moon by the solar panels and 
flexible aluminized mylar/kapton insulation, as shown in Figure 1.1.
The satellite has one mode of operation while taking scientific data. 
It spins and precesses at constant insolation, continuously measuring the
microwave sky.

The instrument measures only temperature differences
from two regions of the sky separated by roughly $140^\circ$.
The instrument is composed of 10 symmetric, passively cooled, 
dual polarization, differential, microwave receivers.
As shown in Table~\ref{tab:inst}, there are four receivers in W band,
two in V band, two in Q band, one in Ka band,
and one in K band. The receivers are fed by two back-to-back
Gregorian telescopes.  

The instrument is passively cooled. The optics, shielded from
Sun and Earth, radiate to free space and cool to 70~K.
Two large (5.6 m$^2$ net) and symmetric radiators
passively cool the front-end microwave electronics
to less than 90~K. One can just make out the heat straps that connect 
the base of the radiators to the microwave components housed below the 
primary reflectors in Figure 1.1. A hexagonal structure, ``hex hub,'' 
between the solar panel array and the 1~m diameter gamma alumina cylinder (GAC)
holds the power supplies, instrument electronics, and 
attitude control systems. The GAC supports a 190~K thermal gradient.
There are no cryogens or mechanical refrigerators and thus
no onboard source of thermal variations. 

\subsection{Optics} The optics comprise two back-to-back shaped
Gregorian telescopes (Barnes et al. 2002; Page \etal~2003a). 
The primary mirrors are 1.4 m $\times$ 1.6 m. The
secondaries are roughly a meter across, though most of the surface 
simply acts as a shield to prevent the feeds from directly viewing the 
Galaxy. The telescopes illuminate 10 scalar feeds on each side, 
a few of which are visible in Figure~\ref{fig:map_line}. 
The primary optical axes are separated by $141^{\circ}$ to allow
differential measurements over large angles on a fast time scale. 
The feed centers occupy a 18 cm $\times$ 20 cm region in the focal plane,
corresponding to a $4^{\circ} \times 4.5^{\circ}$ array on 
the sky, as shown in Figure~\ref{fig:focal} 

At the base of each feed is an orthomode transducer (OMT) that sends the 
two polarizations supported by the feed to separate receiver chains.
The microwave plumbing is such that a single receiver chain 
(half of a ``differencing assembly'') differences electric fields with
two nearly parallel linear polarization vectors, one from each telescope.

Precise knowledge of the beams is essential for accurately computing the 
CMB angular spectrum. The beams are mapped in flight
with the spacecraft in the same observing mode as for 
CMB observations (Page~\etal~2003b). Using Jupiter as a source, we measure
the beam to less than $-30~$dB of its peak value.
Because of the large focal
plane the beams are not symmetric, as shown in Figure~\ref{fig:focal}; 
neither are they Gaussian. In addition, as anticipated,
cool-down distortions of the primary mirrors distort the W-band
and V-band beam shapes. Fortunately, the scan strategy 
effectively symmetrizes the 
beam, greatly facilitating the analysis.

It is as important to understand the sidelobes, as it is to understand
the main beams. We use three methods to assess them. (1)
With physical optics codes\footnote{We use a modified version of the 
DADRA code from YRS Associates (rahmat@ee.ucla.edu).} 
we compute the sidelobe pattern over the full sky and
compute the current distributions on the optics. (2) We built a
specialized test range to make sure that, by measurement, we could 
limit the Sun as a source of spurious signal to $<1~\mu$K level in all
bands. This requires knowing the beam profiles down to 
roughly $-45$ dB (gain above 
%XXX not DBi, right?
isotropic) or $-105$ dB from the W-band peak. We find that over much of
the sky, the measured profiles differ from the predictions at the $-50$~dB
level due to scattering off of the feed horns and the structure that 
holds them. (d) During the early part of the mission, we used
the Moon as a source to verify the ground-based measurements and 
models.
 
With a combination of the models and measurements, we assess both the
polarized and unpolarized  Galactic pickup in the 
sidelobes (Barnes~\etal~2003). In Q, V, and W bands, the
rms Galactic pickup through the sidelobes (with the main beams
at $|b|>15^{\circ}$) is  
$2~\mu$K, $0.3~\mu$K, $0.5~\mu$K, respectively, {\it before 
any modeling}.
As this adds in quadrature to the CMB signal, the effect on the
angular power spectrum is negligible. In K band, because of its low
frequency and large sidelobes, the maps are corrected by 4\% for
the sidelobe contribution. The polarized pickup, which comes mostly
from the passband mismatch, is $\ll 1~\mu$K, except in K band
where it is $\sim 5~\mu$K.

\subsection{Receivers}

The {\sl WMAP} mission was made possible by the HEMT-based amplifiers developed
by Marian Pospieszalski (1992) at the National Radio Astronomy Observatory
(NRAO).  These amplifiers achieve noise
temperatures of 25--100 K at 80~K physical temperature (Pospieszalski \&
Wollack 1998). Of equal importance is that the amplifiers 
can be phase matched 
over a 20\% fractional bandwidth, as is required by the receiver design.
{\sl WMAP} uses a type of correlation receiver to measure the difference in
power coming from the outputs of the OMTs at the base of the feeds
(Jarosik~\etal~2003a). 
All receiver were fully characterized
before flight. The flight sensitivity was within 20\% of expectations. 

\begin{figure*}[t]
\includegraphics[width=1.0\columnwidth,angle=0,clip]{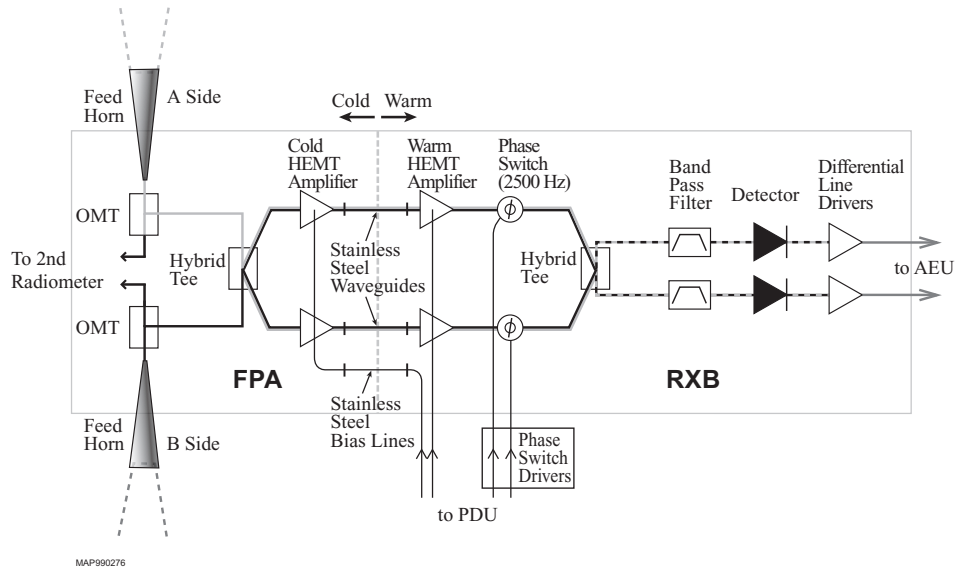}
\vskip 0pt \caption{
One half of a differencing assembly (DA) for detecting
one polarization component. A hybrid Tee (``magic tee'') splits the
inputs into two arms, where the signal is amplified and phase modulated
before recombining. There are two stages of amplification, one 
at 90~K and one at 290~K. The recombined signal is then filtered and
square law detected. The differential output that goes into the
map-making algorithm is derived from the difference of the outputs
of the two line drivers. From Jarosik~\etal~(2003a).
\label{fig:rad}}
\end{figure*}

Figure~\ref{fig:rad} shows the signal path. From the output of the
polarization selector (OMT),
radiation from the two feeds is combined by a hybrid Tee 
into $(A+B)/\sqrt{2}$ and $(A-B)/\sqrt{2}$ signals,
where A and B refer to the amplitudes of the electric fields from one
linear polarization of each feed horn. In one arm of the
receiver, both A and B signals are amplified first by cold ($90~$K) and
then by warm (290 K) HEMT amplifiers. Noise power from the amplifiers,
which far exceeds the input power, is added to each signal by the first 
amplifiers. Ignoring the phase switch for 
a moment, the two arms are then recombined in a second hybrid and both 
outputs of the hybrid are detected after a band-defining filter.
Thus, for each differencing assembly,
there are four detector outputs, two for each polarization.
In a perfectly balanced system, one
detector continuously measures the power in the A signal plus the average
radiometer noise, while the other continuously measures the power in the
B signal plus the average radiometer noise.
 
If the voltage amplification factors for the fields 
are $G_1$ and $G_2$ for the two
arms, the difference in detector outputs is $G_1G_2(A^2-B^2)$. Note that
the average power signal present in both detectors has canceled so that
small gain variations, which have a ``$1/f$'' spectrum\footnote{We use
$f$ for audio frequencies ($<20~$kHz) and $\nu$ for microwave frequencies.}, 
act on the 
difference in powers from the two arms, which corresponds to 
0--3~K, rather than on the total power, which corresponds to roughly 100~K in
the W band. The system is stabilized further by toggling one of the phase switches
at 2.5 kHz and coherently demodulating the detector outputs. 
(The phase switch in the other arm is required to preserve the phase
match between arms and is jammed in one state.) The 2.5 kHz modulation
places the desired signal at a frequency above the $1/f$ knee
of the detectors and audio amplifiers, as well as rejecting any residual
effects due to $1/f$ fluctuations in the gain of the HEMT amplifiers.
The power output of
each detector is averaged for between 51 and 128 ms and telemetered
to the Earth. In total, there are 40 signals (two for each radiometer) 
plus instrument housekeeping data, 
resulting in a data rate of 110~MBy/day.

It is difficult to overemphasize the importance of understanding the
radiometer noise and the stability of the satellite.
One of the requirements for producing high-quality maps
is stable receivers. One aims to measure the largest and smallest
angular scales before the instrument can drift. The main challenge is to
design a system that is stable over the spin rate, when the largest angular
scales are measured. The drift away from stability is generically
termed ``$1/f$''; the effects are similar to $1/f$ noise in amplifiers.  
The {\sl WMAP} noise power spectrum is nearly
white between 0.008 Hz, the spin rate of the 
satellite, and 2.5~kHz; however, there is a slight amount of $1/f$. 
With the exception of the W4 DA, the $1/f$ knee is below 9~mHz.
For W4 it is 26 and 47 mHz for the two polarizations.
To account for this the data are prewhitened 
in the map-making process. This is the primary correction made to the data
and affects only the maps and not the power spectra, as 
discussed below.
 
Although {\sl WMAP}'s differential design was driven by the $1/f$ noise in
the amplifiers, it is also very effective at reducing the effects of
$1/f$ thermal fluctuations of the spacecraft itself. The thermal stability 
of deep space combined with the insensitivity to the spacecraft's
slow temperature variations results in an extremely stable 
instrument. The measured peak-to-peak variation in the temperature
of the cold end of the radiometers at the spin rate 
is $<12~\mu$K. Using the measured radiometer gain susceptibilities, 
this translates to a $<20~$nK radiometric signal. 
If one subtracts the sky signal from the data,
the residual peak-to-peak radiometer output is measured to be
$<0.17~\mu$K at the spin rate. Using the sky-subtracted signal, one
can show that the receiver noise is Gaussian over 5
orders of magnitude (Jarosik~\etal~2003). 
It is because of the stability, and our ability to verify it,
that we can continuously average data together to probe
cosmic signals at the sub-$\mu$K level.  
 
\subsection{Scan Strategy}

High-quality maps are characterized by their fidelity to the 
true sky and their noise properties. Striping, the bane of the map-maker,
results from scans of the sky in which the radiometer output 
departs from the average value for a significant part of the scan. Such
departures, which can be produced by a number of mechanisms, 
are quantified as correlations between pixels. 

A highly interlocking scan strategy is essential for producing a
map with minimal striping. In any measurement, a baseline 
instrumental offset, along with
its associated drift, must be subtracted. Without cross-hatched scans 
this subtraction can preferentially correlate pixels in large swaths, 
resulting in striped maps and substantially more involved 
analyses. {\sl WMAP's} noise matrix is nearly diagonal: 
the correlations between pixels are small. A typical
off-diagonal element of the pixel-pixel covariance matrix
is $<0.1$\% the diagonal value,
except in W4, where it is $\sim0.5$\% at small lag 
(Hinshaw~\etal~2003a).

\begin{figure*}[t]
\includegraphics[width=1.0\columnwidth,angle=0,clip]{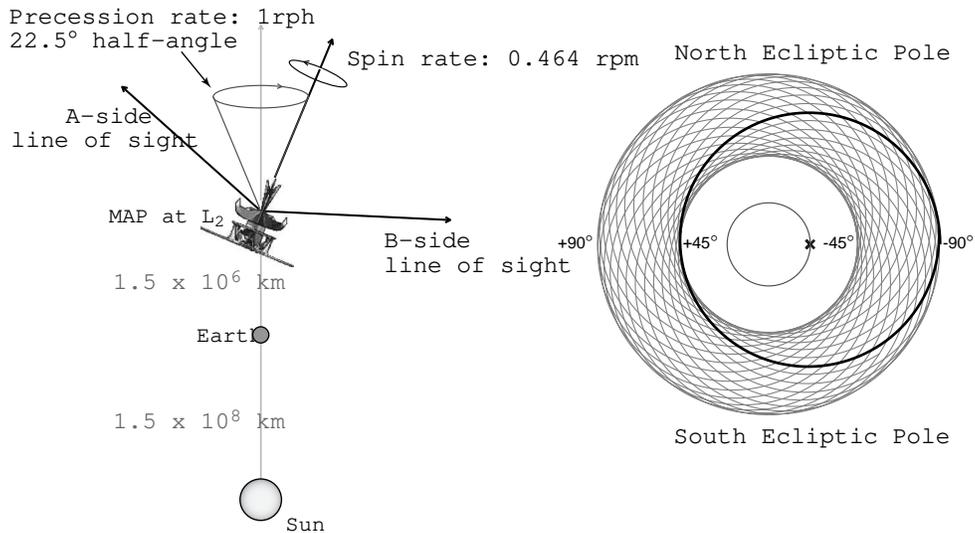}
\vskip 0pt \caption{
{\sl WMAP's} scan pattern from $L_2$. The dark circle on the left-hand
drawing depicts the path covered by two beams for one rotation; the
innermost circle is the path of the spin axis during one precession.
The orbit follows a low-maintenance Lissajous pattern with the Sun,
Earth, and Moon always behind. Corrections to the orbit are made 
roughly every three months. The scan motion is accomplished with three
spinning momentum wheels so that the net angular momentum of the satellite is
near zero. From Bennett~\etal~(2003a).
\label{fig:scan}}
\end{figure*}

To achieve the interlocking scan, {\sl WMAP} spins around its axis 
with a period of 2 min and precesses around a 22\ddeg5 
cone every hour so that the beams
follow a spirograph pattern, as shown in Figure~\ref{fig:scan}. 
Consequently, $\sim 30$\% of the sky 
is covered in one hour, before the instrument temperature 
can change appreciably. The axis of this combined rotation/precession sweeps 
out approximately a great circle as the Earth orbits the Sun. 
In six months, the whole sky is mapped. 
The combination of {\sl WMAP}'s four observing time scales (2.5 kHz, 2.1 min,
1 hour, 6 months) and the heavily interlocked pattern results in a strong
spatial-temporal filter for any signal fixed in the sky. 

Systematic effects at the spin 
period of the satellite are the most difficult to separate from true 
sky signal. Such effects, driven by the Sun, are minimized
because the instrument is always in the 
shadow of the solar array and the precession axis is fixed with
respect to the Sun-{\it WMAP}\ line. Spin-synchronous effects are negligible;
there is no correction for them in the data processing.

With this scan, the instrument is continuously calibrated on the 
CMB dipole. The dominant signal in the 
timeline, the CMB dipole, averages to zero over the 1~hour precession period, 
enabling a clean separation between gain and baseline variations.
The final absolute calibration is actually based on the component 
of the dipole due to the Earth's velocity around the Sun.
As a result, {\sl WMAP} is calibrated to 0.5\% (which will improve).

\section{Basic Results}

The primary scientific result from {\sl WMAP} is a set of maps
of the microwave sky of unprecedented precision and accuracy.
The maps have a well-defined systematic error budget (Hinshaw~\etal~2003a)
and may be used to address the most basic
cosmological questions, as well as to understand Galactic 
and extragalactic emission (Bennett~\etal~2003).

The process of going from the differential radiometer
data to the maps is described in Hinshaw~\etal~(2003a) and 
Wright, Hinshaw, \& Bennett (1996).
Each step of the process is checked with simulations 
that account for all known effects in the data stream.

There are many components of the map-making process that
must be included to produce maps with minimal
systematic error. They include masking bright sources when they are
in one beam, deleting data around the 21 glitches, and solving for the
baseline drift and gain. These effects would
be part of any pipeline. In all,
99\% of the raw data goes directly into the maps.

There are three systematic errors, all of which were anticipated, 
for which corrections are made. They are
(1) the 4\% sidelobe correction in K band that is applied directly
to the K-band map, (2) the $\sim1$\% difference
in loss between the A and B sides, and (3) the prewhitening described above.
The last two are determined with flight data and are applied 
to the time stream. By any standards, this constitutes a remarkably
little amount of correction. Artifacts from the gain, baseline,
and pointing solutions are negligible. 
 
\begin{figure*}[t]
\centering
\includegraphics[width=12cm,angle=0,clip]{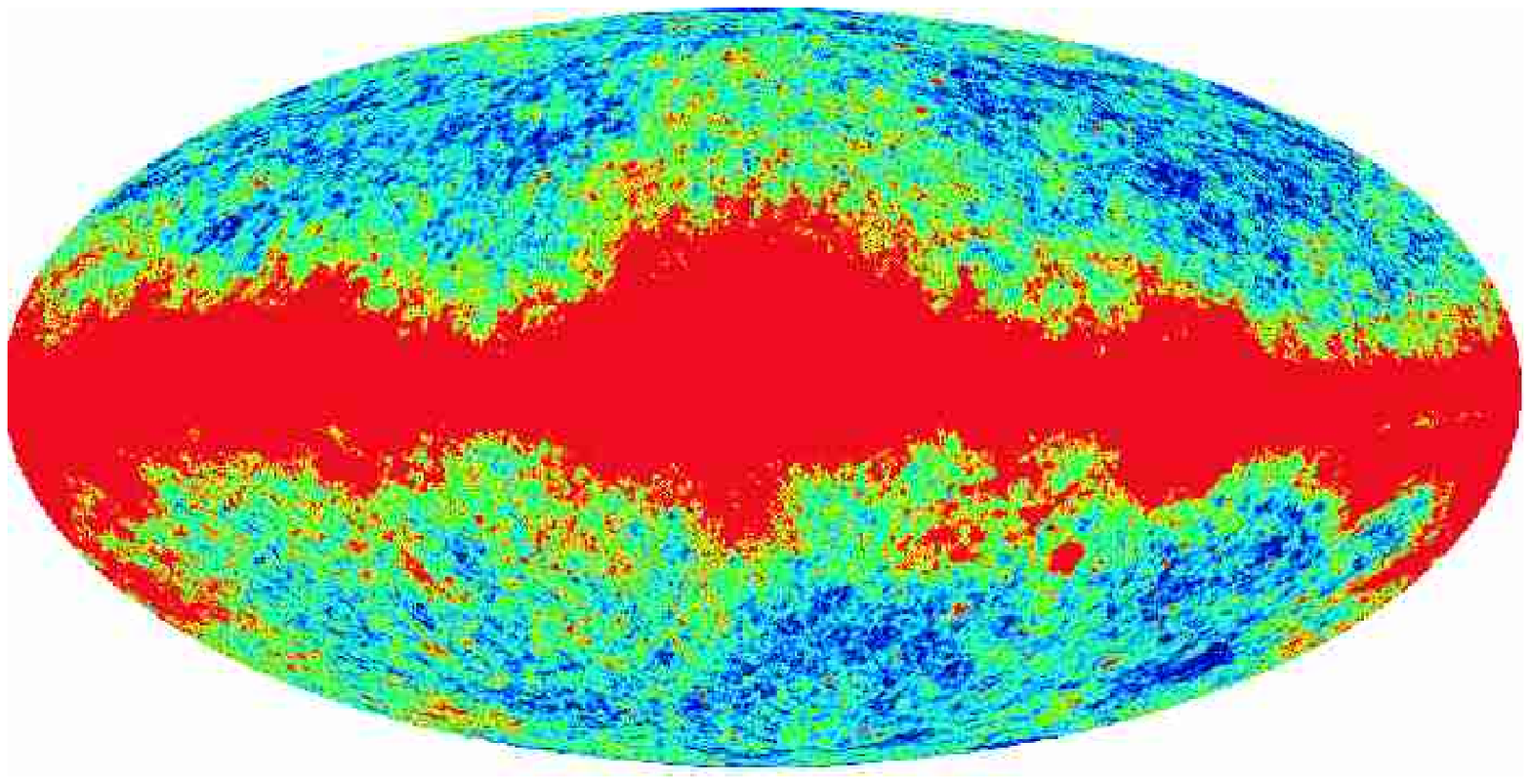}
\includegraphics[width=12cm,angle=0,clip]{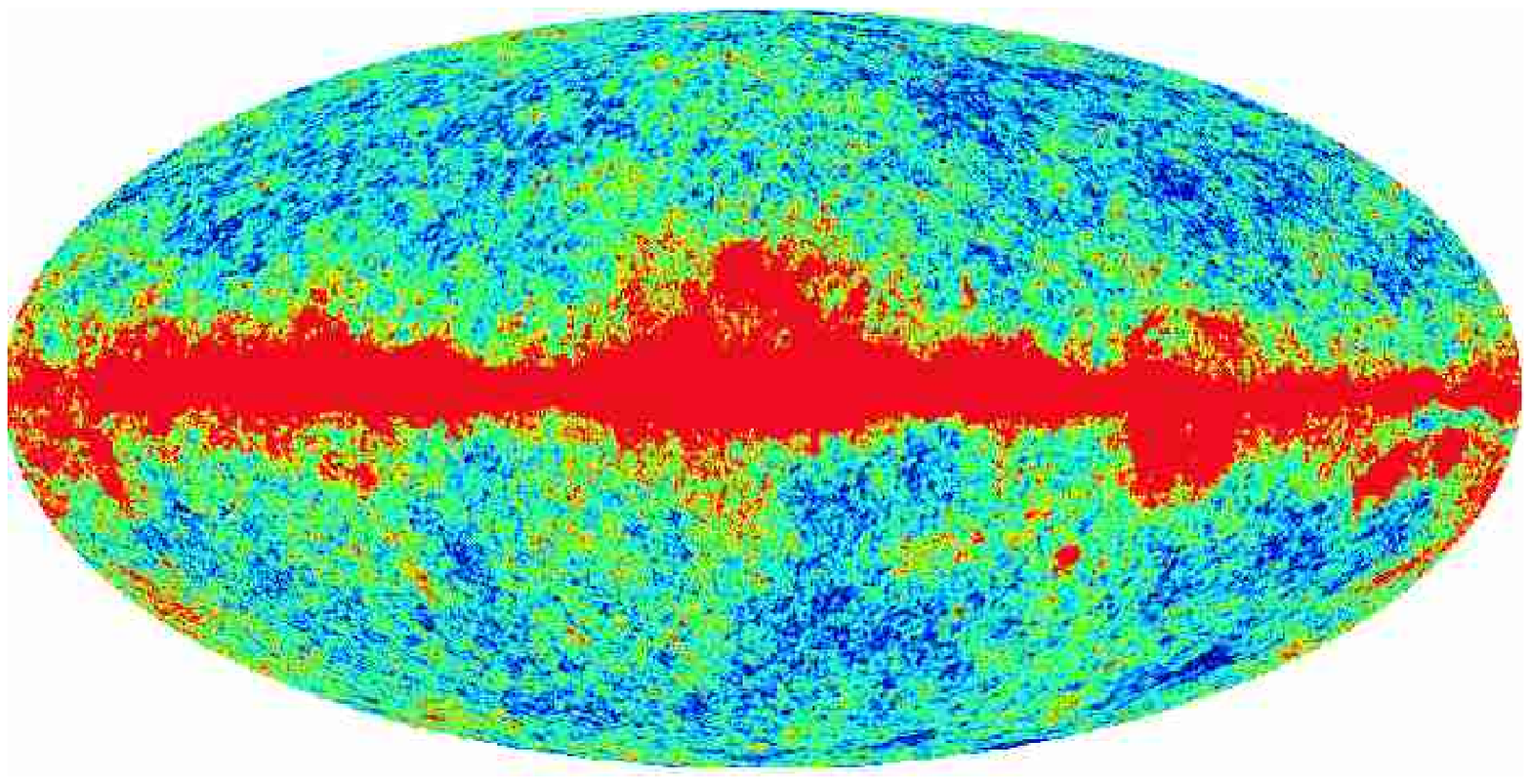}
\includegraphics[width=12cm,angle=0,clip]{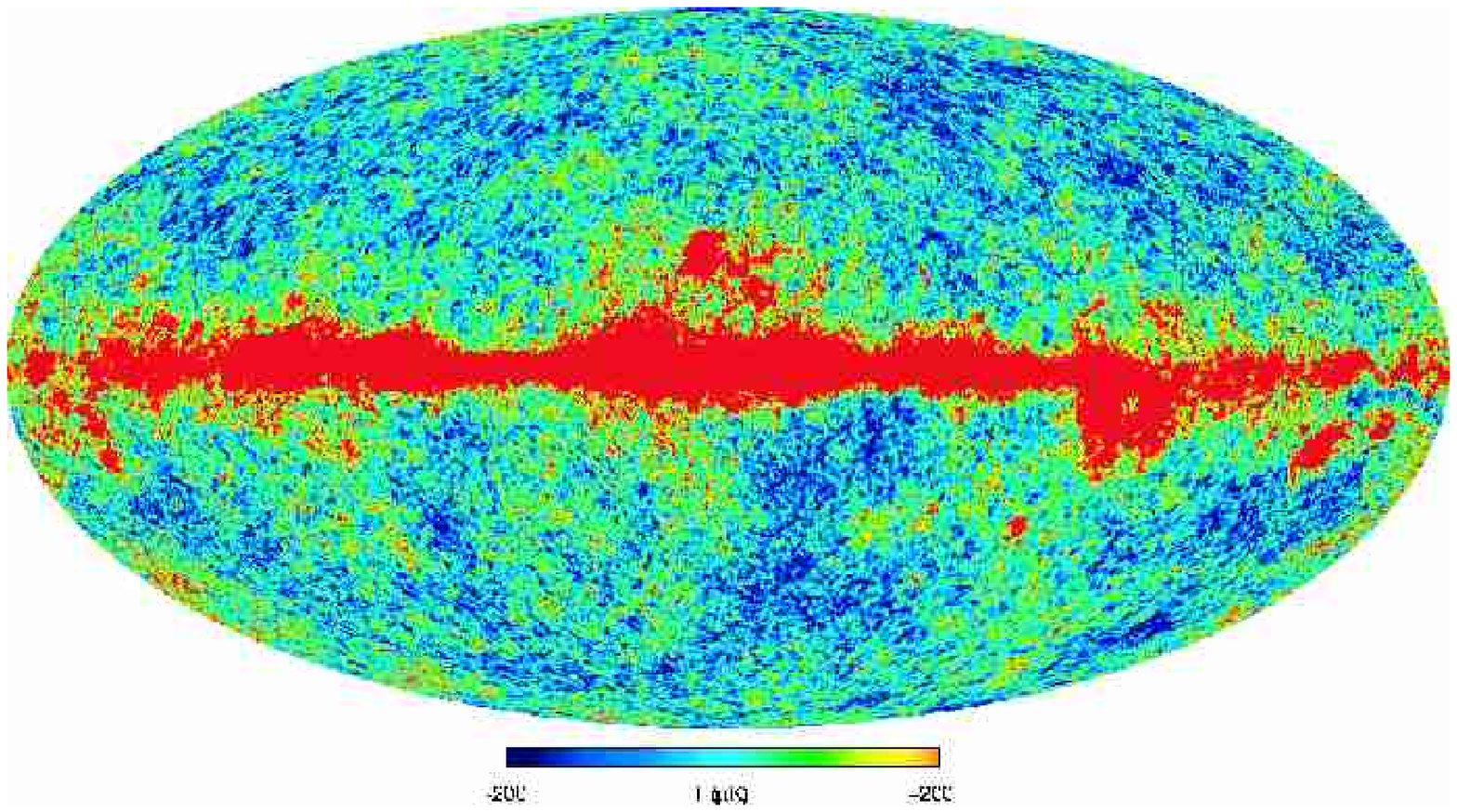}
\vskip 0pt \caption{
K-band (top), Ka-band (middle), and Q-band (bottom) intensity maps in Galactic 
coordinates shown in a Mollweide projection.  The Galactic Center is in 
the middle of each map.  All maps are in CMB thermodynamic units.
From Bennett~\etal~(2003b).
\label{fig:kkaq}}
\end{figure*}

\begin{figure*}[t]
\includegraphics[width=12cm,angle=0,clip]{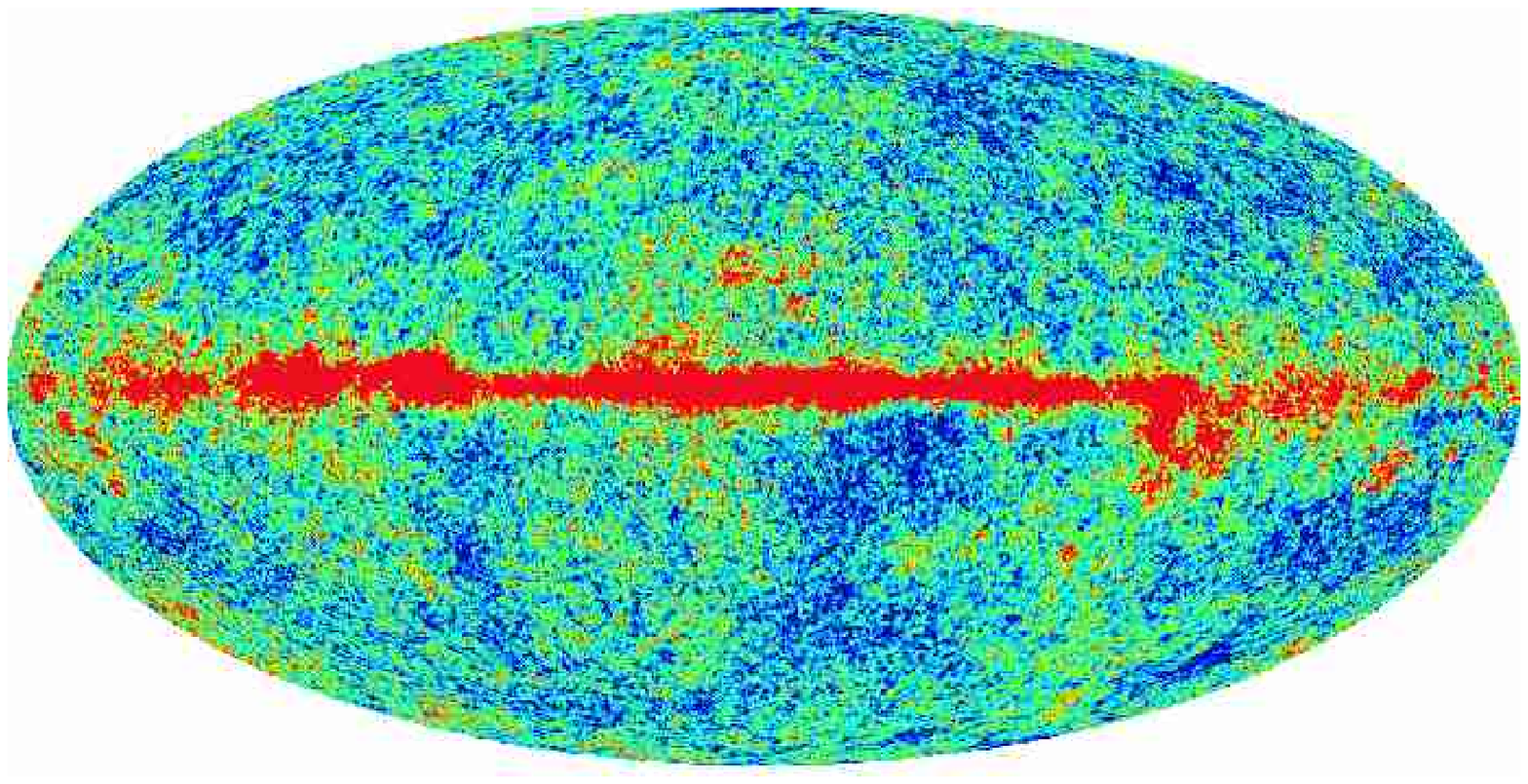}
\includegraphics[width=12cm,angle=0,clip]{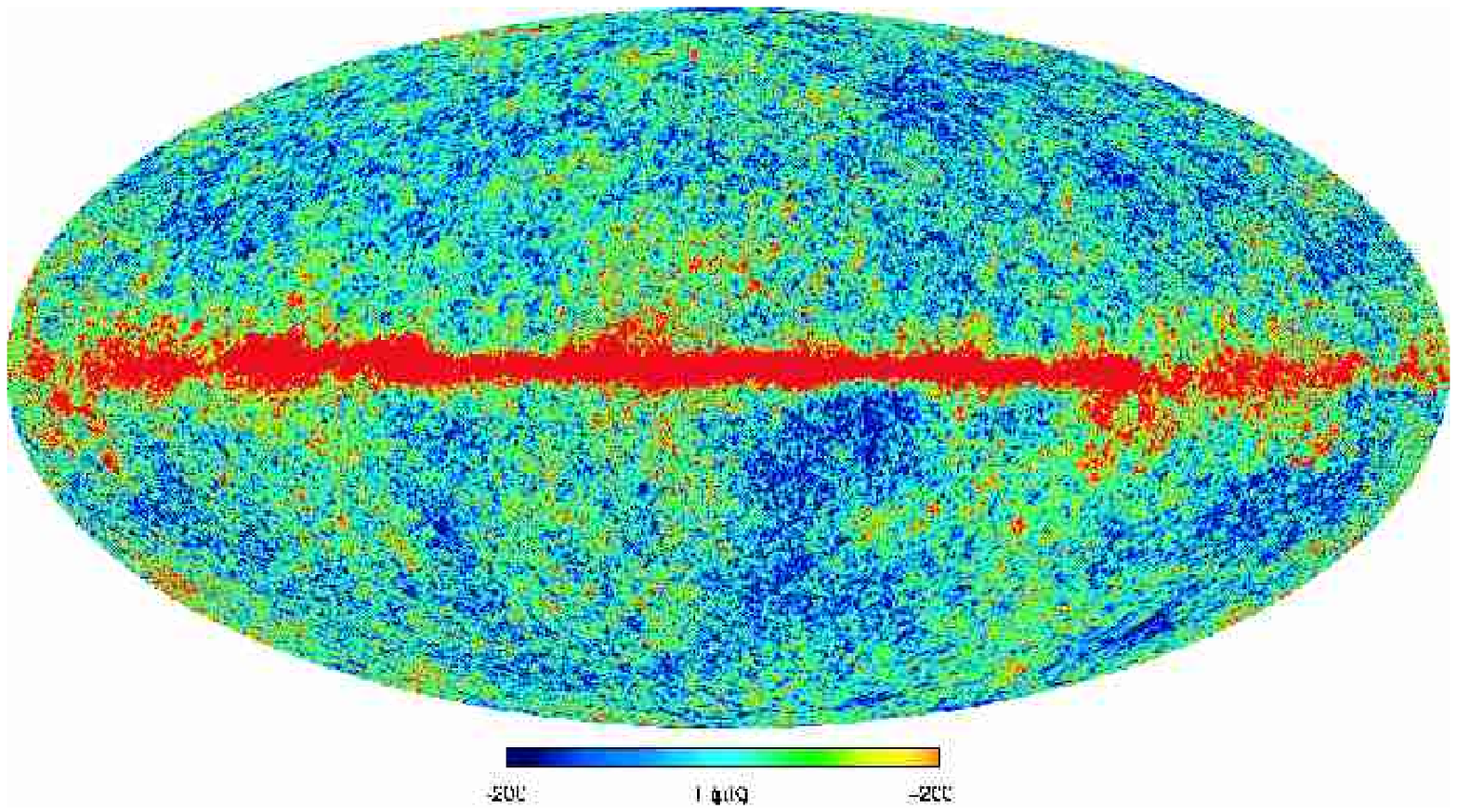}
\vskip 0pt \caption{
V-band (top) and W-band (bottom) maps in the same 
coordinates and scale as the previous three maps.
\label{fig:vw}}
\end{figure*}

Figures~\ref{fig:kkaq} and \ref{fig:vw} show the maps in K through W bands.
All maps have been smoothed to one degree. They are all in 
thermodynamic units relative to a 2.725~K blackbody, so that
the CMB anisotropy has the same contrast in all maps. 
Emission from the Galactic plane saturates the color scale at
all frequencies. At the
lowest frequency, Galactic emission extends quite far off 
the plane. Still, even at 23 GHz,
the anisotropy is clearly evident at high Galactic latitude. 
As one moves to higher frequency, there is less
Galactic emission. When averaged over regions of high Galactic
latitude, foreground contamination is minimum in V band.

\begin{figure*}[t]
\includegraphics[width=14cm,angle=0,clip]{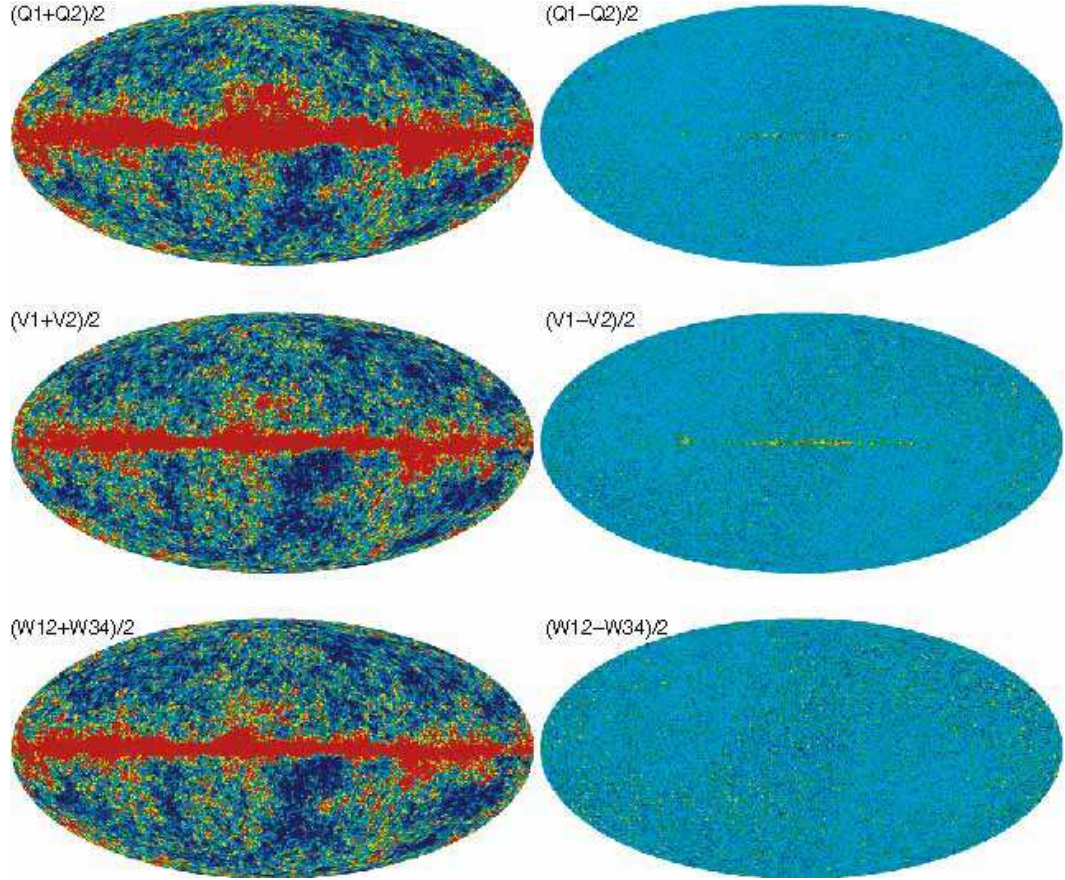}
\vskip 0pt \caption{
This plot shows the sum (left) and difference (right)
maps for Q, V, and W, respectively. A slight bit of contamination 
is evident in V1--V2. This arises because the finite bandwidth
of the passbands. The maps are calibrated 
with the CMB dipole, which has a different spectrum than the
Galactic emission. In the bottom right panel especially,
one can see that the noise in the maps is larger
in the ecliptic plane. This is because the scan pattern
preferentially samples the ecliptic poles. From Hinshaw~\etal~(2003b).
\label{fig:sumdiff}}
\end{figure*}

There are a number checks that can be made of the maps.
The simplest is shown in Figure~\ref{fig:sumdiff}. The sum 
of two channels of the same frequency gives a robust signal
whereas the difference gives almost no signal. With the manifold
instrumental redundancy, a myriad of other internal consistency
checks is possible. We can also compare to the {\sl COBE}/DMR map
and see that, to within the limits
of the noise, {\sl WMAP} and DMR are the same. 
This is a wonderful confirmation of both satellites: they observe
from different places, use different techniques, and have different
systematic errors. 

\subsection{Analysis of Maps}
\label{sec:anal}

The first step in going from the maps to cosmology
is to select a region of sky for analysis. We mask 
regions that are significantly contaminated by diffuse emission
from our Galaxy and by point source emission.
Our selection process is described in Bennett~\etal~(2003c). 

The diffuse contribution comes from free-free emission, synchrotron
emission, and dust emission. To find the most contaminated
sections, one masks regions that are above a threshold in the K-band maps.
The nominal cut excludes 15\% of the sky in a contiguous region near
the Galactic plane. 

One of the surprises from {\sl WMAP} was that a 
successful model of the foregrounds could be made that did
not include a spinning dust component (Draine \& Lazarian 1998). 
Rather, we find that
there is a strong correlation between synchrotron emission and dust, 
and that the synchrotron spectral index varies significantly
over the Galactic plane. Surprising to many, the
synchrotron emission is better correlated with the 
Finkbeiner, Davis, \& Schlegel (1999) dust
map derived from the {\it COBE}/DIRBE and {\it IRAS} at $\nu>200~$GHz
than with the Haslam et al. (1982) synchrotron map at $\nu=0.41~$GHz.

In addition to diffuse emission, we mask a 
0\ddeg6 circle around 700 bright sources that might 
contaminate our results. The sources are selected based on radio 
catalogs, as discussed in Bennett~\etal~(2003c). No correlation with 
infrared sources was found. Contamination by point sources is
a potential problem because their angular power 
spectrum has constant $C_l$, and thus rises as $l^2$ on the
standard plot of the anisotropy. Our approach to identifying
sources is three-pronged: (1) we search the data with a matched filter 
for sources, (2) we marginalize over a point source contribution
in the determination of the angular power spectrum, and (3) we
use a version of the bispectrum tuned to find point sources
to directly 
\clearpage

\clearpage

\noindent
determine their contribution in a map. All three
methods give consistent results when the sources are left in the map
(Komatsu~\etal~2003). All methods indicate that with our masking 
the sources do not contribute to the power spectrum. 
 
After applying the diffuse and point source masks, the remaining
parts of the maps are simultaneously fit to templates of the Finkbeiner et al. 
(1999) dust emission, an H$\alpha$ map that has been correct 
for extinction that traces free-free emission (see Finkbeiner 2003,
and references therein), and a synchrotron map
extrapolated from the 408~MHz Haslam et al. map. 

The template fit removes any remaining residual Galaxy emission, even though 
the fit coefficients do not correspond to the best Galactic model.
Tests are made to ensure that the masking and fitting do not 
bias the results. 
 
With the maps in hand, we show that the fluctuations in the CMB are Gaussian
to the level that can be probed with {\sl WMAP} (Komatsu~\etal~2003). 
This is a key step in the interpretation of the data.
It means that all of the information in the CMB is contained in the angular
power spectrum. In the parlance of CMB analyses, it means that
if the temperature distribution is expanded in spherical harmonics as
\begin{equation}
T(\theta,\phi) = \sum_{lm}a_{lm}Y_{lm}(\theta,\phi),
\end{equation}
that the real and the imaginary parts of the 
$a_{lm}$ are normally distributed. In other words, the phases
between all the harmonics are random. There are no features in the 
CMB like strings or textures that require some definite relation
between the phases. The Gaussanity of the CMB is a triumph
for models such as inflation. However, the degree of non-Gaussanity
expected from these models is a factor of 1000 or more below
where {\sl WMAP} can probe (Maldacena 2002). 

\begin{figure*}[t]
\includegraphics[width=1.0\columnwidth,angle=0,clip]{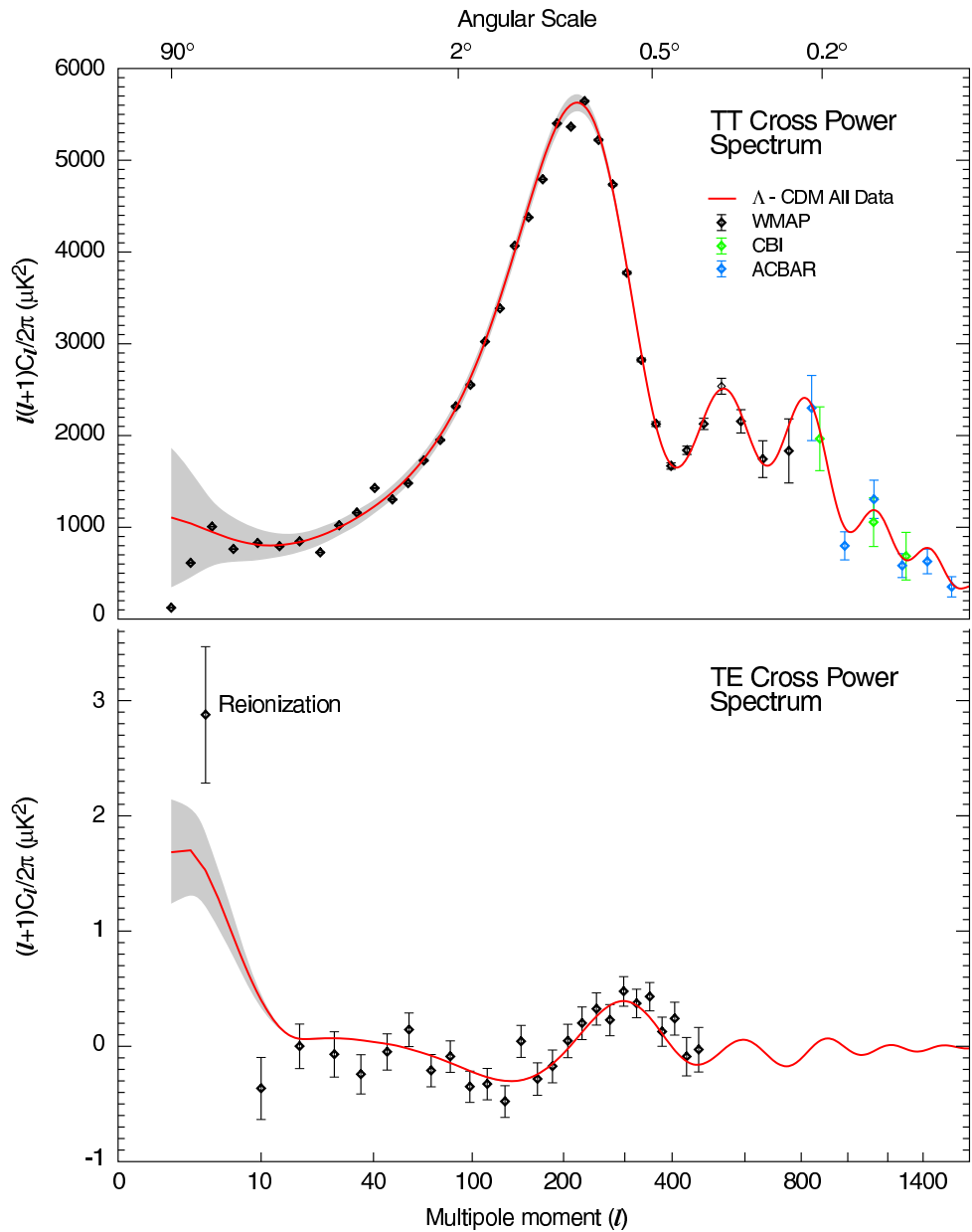}
\vskip 0pt \caption{
The top panel shows the net temperature-temperature
(TT) cross-power spectrum between all channels. The measurement
error bars are shown; the grey band shows the cosmic variance limit.
{\sl WMAP} is cosmic variance dominated up to roughly $\ell=350$.
Below $\ell=100$,
the noise is completely cosmic variance dominated and so we use only
the V and W bands. For $\ell>700$ we augment {\sl WMAP} with the data from
CBI (Mason~\etal~2003) and ACBAR (Kuo~\etal~2003).
The bottom panel shows the
temperature-polarization (TE) cross-power spectrum. The point
labeled ``Reionization'' is the weighted average of the
lowest 8 multipoles. The red line is the best-fit model
as discussed in \S\ref{sec:anal}.
Note that the y-axis of the TE data is scaled to emphasize
the low-$\ell$ region of the spectrum. (Hinshaw~\etal~2003a)
\label{fig:psplot}}
\end{figure*}

From the masked and cleaned maps, we produce the angular
power spectrum, as detailed in Hinshaw~\etal~(2003). All known effects
are taken into account in going from the maps to the spectrum.
We account for the uncertainty in the beam profile, foreground masking, 
and uneven weighting of the sky. Monte Carlo simulations are
used to check that the noise is handled correctly. The stability
of the instrument permits us to determine the noise contribution 
to the percent level.
In producing the power spectrum we use only the cross correlations
between the eight individual Q, V, and W intensity maps that have had the two
polarizations combined (Hinshaw~\etal~2003). 
Because the noise in the maps is, for all intents and purposes, 
independent, there are no effects from prewhitening or from
the peculiarities of any one DA in the power spectrum. Of the possible
36 separate measures of the power spectrum, we use 28.  
The results are shown in the top panel of Figure~\ref{fig:psplot}.
  
We also produce maps of the Stokes $Q$ and $U$ polarization components
(Hinshaw~\etal~2003a; Kogut~\etal~2003).
These maps are produced from the difference
of the two differential measurements common to one pair of feeds
and use lines of constant Galactic longitude as a reference direction.
Though the polarization maps pass a series of null tests, there
was not enough time to prepare them for the first release.
However, the cross correlation between the polarization and 
temperature maps is much less susceptible to systematic error than 
the polarization maps alone and may be more easily determined with confidence.
At $>10\,\sigma$, we find a correlation between the Stokes $I$ (temperature)
and $Q$ maps. This is due to the $E$ mode of the polarization
(Kamionkowski, Kosowsky, \& Stebbins 1997; Zaldarriaga \& Seljak 1997). 
In Q, V, and W bands, the TE (Stokes $I$ cross $Q$) 
has a thermal frequency spectrum; it
is not due to foregrounds. The correlation between Stokes $I$ and $U$, 
which is significant for $B$ modes, is consistent with zero.  

The TE cross-power spectrum is shown in 
the bottom panel of Figure~\ref{fig:psplot}. We see that there is a 
model that describes the TT and TE data beautifully. As $E$ modes
at $1^\circ$ scales are generated only during the transition 
from an ionized to a neutral Universe,
the agreement indicates that we understand the physics of the 
decoupling process. Insofar as the  $\ell>10$
TE correlation is predicted from the TT data, it does
not contribute to our understanding of the cosmological parameters.

The TE correlation for $\ell<10$, however,
considerably changes our understanding
of cosmic history. It was a surprising discovery. 
The only way to polarize the CMB at these large angular
scales is through reionization\footnote{Tensor modes can generate 
an $\ell<10$ TE signal, but the required tensor amplitude is inconsistent
with the TT data.}. In our fiducial model, we assume that the first
generation of stars completely reionized the Universe at a fixed 
redshift, $z_r$. The optical depth, $\tau$, is then just the line integral
out to $z_r$. We find that $\tau=0.17\pm0.04$.
Because quasar spectra show there is neutral hydrogen at $z=6$
(Becker~\etal~2001; Djorgovski~\etal~2001; Fan~\etal~2002),
the ionization history may not be as simple as in our model. After accounting
for uncertainties in our model, we find
$z_r=20^{+10}_{-9}$ (Kogut~\etal~2003).

The large optical depth changes the way one interprets the
TT power spectrum. It means that the intrinsic CMB fluctuations
are 30\% larger than the TT plot shows. This in turn affects
the $\sigma_8$ (or overall normalization) one gets for the CMB
data. 

It is the maps and the power spectra that we derive from them
that have the lasting value. 

\clearpage
\clearpage

\noindent
Producing them is the primary goal 
of the mission. The maps do not depend on a particular
cosmological model, the temperature power spectrum 
does not depend on a model, and the polarization does not depend on a model. 
We next focus more on to the cosmological interpretation of the data.

\begin{figure*}[t]
\includegraphics[width=1.0\columnwidth,angle=0,clip]{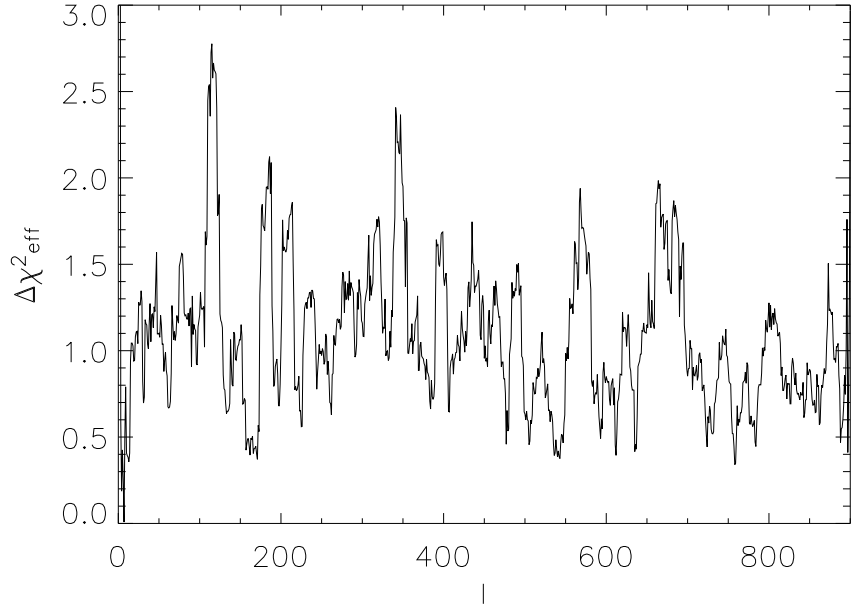}
\vskip 0pt \caption{
The contribution to the $\chi^2$ as a function of $\ell$.
The spikes in this plot may be identified with departures from
the best-fit model in Figure~\ref{fig:psplot}. The binning here is
per $\ell$, whereas  Figure~\ref{fig:psplot} has values binned together.
From Spergel~\etal~2003.
\label{fig:chisq}}
\end{figure*}

\subsection{Cosmic Parameters from Maps}

One can deduce the parameters of the cosmological model
that fit the data. The procedure is conceptually straightforward,
though the devil is in the details. Advances
have been made since the first estimates. One now uses
numerically friendly combinations of parameters (e.g., 
Kosowsky, Milosavljevi\'c, \& Jimenez  2002), customized versions
of CMBFAST-like programs (Seljak \& Zaldarriaga 1996), 
and Markov Chain Monte Carlos (Christensen~\etal~2001).

For any model fitting, one assumes some prior information. For example,
one might limit oneself to only adiabatic $\Lambda$CDM models, or,
more restrictively, to models with $h>0.5$, or flat models, or
models that agree with large-scale structure data. A goal is to say
as much as possible with the minimum number of priors.

For the simplest models of the CMB, there is an intrinsic parameter
degeneracy called the ``geometric degeneracy'' (Bond~\etal~1994;
Zaldarriaga, Spergel, \& Seljak 1997).
Using the TT CMB data alone, one cannot separately determine
$\Omega_m$, $h$, and $\Omega_\Lambda$ even with cosmic variance
limited data out to $\ell=2000$. One can play these
parameters off one another to produce identical spectra. 
The degeneracy is broken by picking a value of the Hubble
constant, assuming a flat geometry, or something similar.
With the {\sl WMAP} data, supernovae data (Riess et al. 1998; 
Perlmutter~\etal~1999), and the {\it HST} Key Project value of 
$h=0.72\pm3\pm7$ (Freedman~\etal~2001), $\Omega_T=1.02\pm0.02$.
With the {\sl WMAP} data and a prior of $h>0.5$,  $0.98<\Omega_T<1.08$
(95\% confidence). While this does not {\it prove} that the Universe is 
geometrically flat, 
Occam's razor and the Dicke arguments that in part inspired 
inflation lead one to take a flat Universe as 
the basic model\footnote{If $\Omega_\Lambda=0$, then just the position
of the first peak shows that the Universe is flat (Kamionkowski,
Spergel, \& Sugiyama  1994).}.
 
Unless otherwise noted, all the parameters quoted have a prior that
the Universe is geometrically flat. With this, the values in 
Table~\ref{tab:params} are obtained. We give the values for
just power-law models for the index. Values for a variety of
models and parameter combinations are given in Spergel~\etal~(2003). 

\begin{table}
%\centering
\caption{Sample of Best-Fit Cosmological Parameters
\label{tab:params}}
\begin{tabular}{c|cccc}
\hline \hline
Parameter & &{\sl WMAP} only, flat & {\sl WMAP}+others & Non-CMB estimates\\   
\hline
      $A$          & & $0.9\pm 0.1$     
                     & $0.75^{+0.08}_{-0.07}$  & ... \\
      $\tau$       & & $0.166^{+0.076}_{-0.071}$ 
                     & $0.117^{+0.057}_{-0.053}$ &...\\
      $\Omega_bh^2$& & $0.024\pm 0.001$ 
                     &$0.0226\pm 0.0008$  & $0.021-0.025$ \\
      $\Omega_mh^2$& & $0.14\pm 0.02$   
                     & $0.133\pm 0.006$ & $0.062-0.18$ \\
      $h$          & & $0.72\pm 0.05$   
                     &$0.72\pm 0.03$  & $h=0.72\pm3\pm7$ \\
      $n_s$        & & $0.99\pm 0.04$   
                     & $0.96\pm 0.02$  & ... \\
      $\chi^2_{eff}/\nu$ && $1431/1342=1.066$  
                   & ...   & ... \\
\hline \hline
\end{tabular}

\footnotesize{The best-fit parameters for the {\sl WMAP} data for two of the
models tested. The second column is for {\sl WMAP} alone; the third is 
for a combination of {\sl WMAP}+CBI+ACBAR+2dFGRS+Ly$\alpha$.
The $\chi^2$ is not given because the Ly$\alpha$ data are correlated. 
The non-CMB estimates are from studies of the Ly$\alpha$ D/H and DLA
systems, cluster number counts, galaxy velocities, and the {\it HST} Key 
Project.  They come from many studies and are discussed in 
Spergel~\etal~(2003).}
\end{table}

\begin{figure*}[t]
\includegraphics[width=0.95\columnwidth,angle=0,clip]{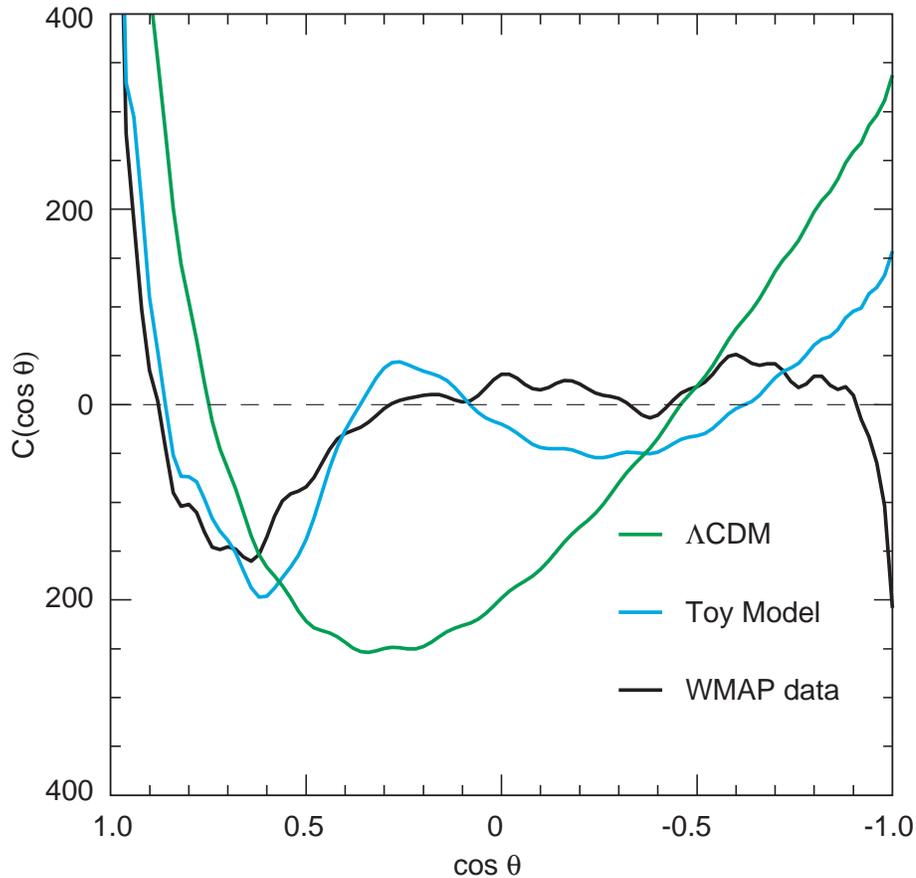}
\vskip 0pt \caption{
Correlation function of the {\sl WMAP} data.
The green line corresponds to a $\Lambda$CDM model. The black line
is the {\sl  WMAP} data. The ``Toy''
model, described in Spergel~\etal~(2003), has been tailored to
fit the data by putting discrete modes in the matter power spectrum.
\label{fig:corrfcn}}
\end{figure*}

It is of interest that the reduced $\chi^2$ for the best-fit model
has a probability to exceed of $\sim 5$\%. The reduced $\chi^2$
%XXX exceed what?
is slightly large, but on its own does not signal that the model is wrong. 
In fact, no matter
what model we try, $\chi^2/\nu$ more or less stays the same.
If isocurvature modes are added to fit the low-$\ell$ region 
of the spectrum,  $\chi^2/\nu=1468/1378=1.065$ (Peiris~\etal~ 2003); 
if one adds tensor modes, allows the spectral index $n_s$
to run, and the 2dFRGS data are added, $\chi^2/\nu=1465/1379=1.062$;
if a step is added to the power spectrum that attempts to fit
bumps and wiggles in the {\sl WMAP} angular power spectrum,  
$\chi^2/\nu=1422/1339=1.062$ (Peiris~\etal~ 2003). 
In other words, the admixture of additional model elements
does not substantially improve the fit over the basic model.
The contributions to $\chi^2/\nu$ are 
shown in Figure~\ref{fig:chisq}. It is likely that, as the sky is covered
more thoroughly, as we understand the beams better, and as we
account for more detailed astrophysical effects (Spergel~\etal~ 2003),
$\chi^2$ will decrease. 

Though the simple model is a wonderful fit to the {\sl WMAP}
data, the small quadrupole and the related lack of correlation
at large angular separations is striking. Though largely obscured 
by cosmic variance, there appears not even a hint of an upturn at low $\ell$,
as expected in $\Lambda$CDM models. 
(However, {\sl COBE} also found no evidence for such an upturn). 
This departure from the model constitutes a small (though possibly
important) fraction of the total fluctuation power and does
not cast doubt on the interpretation of the $\ell>40$ spectrum.
The correlation function is shown in Figure~\ref{fig:corrfcn}.
For $\theta>60^{\circ}$
the probability that the {\sl WMAP} data agree with the best-fit 
$\Lambda$CDM models {\it at these angular scales} is $\sim 1/500$.
One must keep in mind that the statistic is {\it a posteriori}, but it is 
still unsettling\footnote{David Spergel was overheard saying, 
``One should not obsess about the low quadrupole. One should
not obsess about the low quadrupole. Really, one should not obsess
about the low quadrupole.''}. 

\section{Summary and the Future }

Before it was widely appreciated that so many of the
cosmological parameters could be deduced directly from the 
CMB (e.g., Jungman~\etal~1995), the motivation for studying 
the anisotropy was 
that it would tell us the initial conditions for the formation 
of cosmic structure and would tell us the cosmogony in which
to interpret that process. In this vein, we summarize here 
some of the grander conclusions we have learned from {\sl WMAP}.

\begin{enumerate}

\item A flat $\Lambda$-dominated CDM model with just six parameters ($A$, 
$\tau$, $\Omega_m$, $\Omega_b$, $h$, and $n_s$) gives an excellent description 
of the statistical properties of $>10^6$ measurements of the anisotropy over 
85\% of the sky. This does not mean that a simple flat $\Lambda$CDM model is 
complete, but any other model must look very much like it. 

\item An Einstein-de~Sitter model, which is flat with $\Omega_\Lambda=0$, 
is ruled out at $>5\,\sigma$. This statement is based on just the CMB,
with no prior information (e.g., no prior on $h$). Stated another way, if 
$h>0.5$ then {\sl WMAP} requires $\Omega_\Lambda\ne0$ in the angular-diameter 
distance to the decoupling surface.

\item A closed model with $\Omega_m=1.28$ and  $\Omega_\Lambda=0$ fits the data
but requires $h=0.33$, in conflict with {\it HST} observations. It also 
disagrees with cluster abundances, velocity flows, and other cosmic probes.

\item The fluctuations in the metric are superhorizon (Peiris~\etal~2003). 
Turok (1996) showed that in principle one can construct models based on 
subhorizon processes that can mimic the TT $\Lambda$CDM spectra. However, 
these models cannot mimic the observed TE anticorrelation at $50<\ell<150$
(Spergel \& Zaldarriaga 1997).
  
\end{enumerate} 

By combining {\sl WMAP} data with other probes we can constrain the mass 
of the neutrino and constrain the equation of state, $w$.  As our knowledge of the 
large-scale structure matures, in particular of the Ly$\alpha$ forest 
(McDonald~\etal~ 2000; Zaldarriaga, Hui, \& Tegmark 2001; Croft~\etal~ 2002; 
Gnedin\& Hamilton~ 2002; Seljak, McDonald, \& Markarov 2003), we can use the 
combination of these measurements to probe in detail the spectral index as 
a function of scale, thereby directly constraining inflation and other related 
models.  No doubt correlations with other measurements will shed light on other
cosmic phenomena. With {\sl WMAP}, physical cosmology has moved from 
assuming a model and deducing cosmic parameters to detailed testing 
of a standard model.

The {\sl WMAP} mission is currently funded to run through October 2005, 
though as of
this writing there is nothing to limit the satellite from operating longer. 
The quality of the maps will continue to improve as the 
mission progresses. Not only will the noise integrate down, but
our knowledge of the beams will continue to improve, and 
we will make corrections that were too small
to consider for the year one analysis. For example,
increased knowledge of the beams will reduce the uncertainties
in the power spectrum at high $\ell$ and will
permit an even cleaner separation of the CMB and foreground
components near the Galactic plane. 
It is the set of high-fidelity maps of the microwave
sky that will be {\sl WMAP}'s most important legacy. 

With four-year maps, the polarization at intermediate 
and large angular scale should be seen and the 
epoch of reionization will be much better constrained. It may
even be possible to detect the lensing of the CMB by the
intervening mass fluctuations.

On the horizon is the {\sl Planck} satellite, which, with its higher 
sensitivity, greater resolution, and larger frequency range should,  
significantly expand upon the picture presented by {\sl WMAP}. In addition,
a host of other measurements that push to yet 
finer angular scales
and a future satellite dedicated to measuring the polarization in the CMB,
especially the $B$ modes, promise a rich and exciting future for CMB studies. 

More information about {\sl WMAP} and the data are available 
through: \\
http://lambda.gsfc.nasa.gov/


\begin{thereferences}{}

\bibitem{barnes02} 
Barnes, C., et al. 2002, \apjs, 143, 567 

\bibitem{barnes03} 
------.  2003, \apj, submitted (astro-ph/0302215)

\bibitem{beck}
Becker, R. H., et al. 2001, \aj, 122, 2850

\bibitem{bennett03a} 
Bennett, C. L., et al.  2003a, \apj, 583, 1 

\bibitem{bennett03b} 
------.  2003b, \apj, submitted (astro-ph/0302207)

\bibitem{bennett03c} 
------.  2003c, \apj, submitted (astro-ph/0302208)

\bibitem{bond94} 
Bond, J. R., Crittenden, R., Davis, R.~L., Efstathiou, G., \& Steinhardt, 
P.~J. 1994, Phys. Rev. Lett., 72, 13

\bibitem{nelson01} 
Christensen, N., Meyer, R., Knox, L., \& Luey, B. 2001, Classical and Quantum 
Gravity, 18, 2677

\bibitem{colless01} 
Colless, M., et al.  2001, \mnras, 328, 1039

\bibitem{croft02} 
Croft, R. A. C., Weinberg, D. H., Bolte, M., Burles, S., Hernquist, L., Katz, 
N., Kirkman, D., \& Tytler, D. 2002, \apj, 581, 20

\bibitem{djetal} 
Djorgovski, S. G., Castro, S., Stern, D., \& Mahabal, A. A. 2001, ApJ, 560, L5

\bibitem{DL98} 
Draine, B. T., \& Lazarian, A. 1998, \apj, 494, L19

\bibitem{Fan} 
Fan, X., et al. 2002, \aj, 123, 1247

\bibitem{Finkbeiner03} 
Finkbeiner, D. P. 2003, \apjs, 146, 407

\bibitem{Finkbeiner99} 
Finkbeiner, D.~P., Davis, M., \& Schlegel, D.~J. 1999, \apj, 524, 867

\bibitem{Freedman01} 
Freedman, W. L.,  et al. 2001, \apj, 553, 47

\bibitem{Gnedin02} 
Gnedin, N. Y., \& Hamilton, A. J. S. 2002 MNRAS, 334, 107

\bibitem{Haslam82}
Haslam, C. G. T., Stoffel, H., Salter, C. J., \& Wilson, W. E. 1982, A\&AS, 
47, 1

\bibitem{hinshaw03a} 
Hinshaw, G., et al.  2003a, \apj, submitted (astro-ph/0302222)

\bibitem{hinshaw03b} 
------.  2003b, \apj, submitted (astro-ph/0302217)

\bibitem{jarosik03a} 
Jarosik, N., et al.  2003a, \apjs, 145, 413

\bibitem{jarosik03b} 
------. 2003b, \apj, submitted (astro-ph/0302224)

\bibitem{jung95} 
Jungman, G., Kamionkowski, M., Kosowsky, A., \& Spergel, D. N. 1995, Phys. 
Rev. D., 54, 1332

\bibitem{kks} 
Kamionkowski, M., Kosowsky, A., \& Stebbins, A. 1997, Phys. Rev. D, 55, 7368

\bibitem{kss} 
Kamionkowski, M.,  Spergel, D.~N., \& Sugiyama, N. 1994, \apj, 426, L57

\bibitem{kogut03} 
Kogut, A., et al.  2003, \apj, submitted (astro-ph/0302213)

\bibitem{komatsu03} 
Komatsu, E., et al.  2003, \apj, submitted (astro-ph/0302223)

\bibitem{arthur02} 
Kosowsky, A., Milosavljevi\'c, M., \& Jimenez, R. 2002, Phys. Rev. D, 66, 63007

\bibitem{kuo02} 
Kuo, C. L., et al. 2003, \apj, submitted (astro-ph/0212289)

\bibitem{maldacena02} 
Maldacena, J. 2002, preprint (astro-ph/0210603)

\bibitem{mason02} 
Mason, B.~S., et al. 2003, \apj, in press (astro-ph/0205384)

\bibitem{} 
McDonald, P., Miralda-Escud\'e, J., Rauch, M., Sargent, W. L. W., Barlow, 
T. A., Cen, R., \& Ostriker, J. P. 2000, \apj, 543, 1

\bibitem{page03a} 
Page, L. A., et al. 2003a, \apj, 585, 566

\bibitem{page03b} 
------. 2003b, \apj, submitted (astro-ph/0302214)

\bibitem{page03c} 
------. 2003c, \apj, submitted (astro-ph/0302220)

\bibitem{peiris03} 
Peiris, H., et al.  2003, \apj, submitted (astro-ph/0302225)

\bibitem{perl99} 
Perlmutter, S., et al. 1999, \apj, 517, 565 

\bibitem{posp92} 
Pospieszalski, M. W. 1992, Proc. IEEE Microwave Theory Tech., MTT-S 1369

\bibitem{} 
------. 1997, Microwave Background Anisotropies, ed.
F. R. Bouchet (Gif-sur-Yvette: Editions Fronti\`ers), 23

\bibitem{pospwol98} 
Pospieszalski, M. W., \& Wollack, E. W.  1998, IEEE Microwave Theory Tech., 
MTT-S Digest, Baltimore, MD 669

\bibitem{riessl98} 
Riess, A.~G., et al. 1998, \aj, 116, 1009 

\bibitem{mcd2000} 
Seljak, U., McDonald, P., \& Markarov, A. 2003, \mnras, in press 
(astro-ph/0302571)

\bibitem{selzal} 
Seljak, U., \& Zaldarriaga, M. 1996, \apj, 469, 437 

\bibitem{spergel03} 
Spergel, D. N., et al.  2003, \apj, submitted (astro-ph/0302209)

\bibitem{sperzal} 
Spergel, D. N., \& Zaldarriaga, M. 1997, Phys. Rev. Lett., 79, 2180

\bibitem{turok96} 
Turok, N. 1996, Phys. Rev. D, 54, 3686

\bibitem{verde03} 
Verde, L., et al. 2003, \apj, submitted (astro-ph/0302218)

\bibitem{wright96} 
Wright, E. L., Hinshaw, G., \& Bennett, C. L. 1996, \apj, 458, L53

\bibitem{zht} 
Zaldarriaga, M., Hui, L., \& Tegmark, M. 2001, \apj, 557, 519 

\bibitem{zalsel} 
Zaldarriaga, M., \& Seljak, U. 1997, Phys. Rev. D, 55, 1830 

\bibitem{zalspersel} 
Zaldarriaga, M., Spergel, D. N., \& Seljak, U. 1997, \apj, 488, 1 

\end{thereferences}
\end{document}